\documentclass{article}[12]
\usepackage{latexsym}
\begin{document} 
\newcommand\Prob{{\bf Prob}}
\newcommand\given{{\, | \,}}
\newcommand\convD{\buildrel {\cal D} \over \longrightarrow }  
\newtheorem{theorem}{Theorem}
\newcommand\E{{\bf E}}
\newcommand\V{{\bf Var}}
\newcommand\indicator[1]{{\bf 1}_#1}
\newcommand\normal{{\cal N}}
\newcommand\doubleint{\mathop{\int\int}}
\newcommand\almostsure{\buildrel {a.s.} \over  \longrightarrow}
\newtheorem{corollary}{Corollary}
\newtheorem{lemma}{Lemma}
\newcommand\proof{{\it Proof}}
\title{ Random Sequential  Generation  of Intervals 
 for the Cascade Model of  Food Webs}

\author{ 
Yoshiaki Itoh \footnote {Email: itoh@ism.ac.jp}  \\
{\small The Institute of Statistical Mathematics and }\\{\small the Graduate University for 
Advanced Studies} 
\\{\small  10-3 Midori-cho, Tachikawa, Tokyo 190-8562, Japan}
}

\date{ June 23, 2011}
\maketitle

\vspace{0.3cm}
\noindent 
{\bf Abstract}  
The cascade model generates  a food web at random. 
In it the species are labeled from $0$ to $m$, and  arcs are given at 
random between pairs of the species. For an arc with endpoints $i$ and $j$ 
($i<j$), the species $i$ is eaten by the  species labeled $j$. 
The chain length (height), generated at random,  models  the length of food chain in ecological data.
The aim of this note is to introduce the random sequential   generation  of 
intervals   as   a Poisson  model  
which  gives naturally an analogous  behavior to the cascade model.

\vspace{.5cm}

\vspace{.5cm}

\newpage
\section{Introduction} 
The cascade model (Cohen (1990), Cohen and Newman (1985, 1986), Cohen, Briand and Newman (1990),  Newman (1992)) 
is introduced to explain the ecological data on community food webs. The model generates  a food web at random.
In it the species are labeled from $0$ to $m$, and  arcs are given at 
random between pairs of the species. For an arc with endpoints $i$ and $j$ 
($i<j$), the species $i$ is eaten by the  species labeled $j$. 
The chain length (height) is   compared with  the length of food chain in ecological data.
 The problem of chain length of food webs gives   natural questions for random evolution of graphs 
in which  connectedness and other related problems are studied (Erd\"os and R\'enyi (1960), Shepp (1989), 
Durrett and Kesten (1990)).

 The random sequential bisection of intervals (Sibuya and Itoh (1987), 
Janson and Neininger (2008), Dutour Sikiric and Itoh (2011) ) gives an analogous asymptotic behavior to the   binary search tree 
(Robson (1979), Flajolet and Odlyzko (1982), Mahmoud and Pittel (1984), Devroye (1986)). 
Here we  introduce  a random sequential  generation of intervals   
to understand  the chain length for 
the cascade model,  extending the idea of  random sequential bisection of intervals.    
We  can naturally obtain a formula  for  the chain length, as given in section 2 and section 3, which  suggests 
 the corresponding formula  for 
the original cascade model as given in section 4. 
The   random sequential interval generation   is   a  Poisson approximation   
 for  the cascade model of food webs, in the  sense given in  equation (\ref{eq:m3}) in section 3, and equation  (\ref{eq:t}) in 
section 4.

Consider the random oriented graph with vertex set $\{0,1,2,...,m\}$ in which the 
${m+1\choose2}$ oriented edges $(i,j)$ with $i<j$ occur independently of each 
other with probability $p=c$ 
and no 
edge $ j \leq i$ occurs. 
Let us give  a random variable $X_{ij}$ to  each oriented edge $(i,j)$ 
with $i<j$ with the vertex set 
$\{0,1,2,...,m\}$, where $X_{ij}$ are  mutually independent   random  variable with $X_{ij}=1$
with probability $p=c$ 
and $X_{ij}=0$ with probability $1-c$.  

 Consider the random variables $X_{ij}$ with $i=0$.   
For sufficiently small $c$, the number $N=\sum_{j=1}^m X_{ij}$ with $i=0$ is approximately 
distributed by the Poisson distribution $\frac{(c\, m)^k}{k!} e^{-c\,m}$. 
Let each of $N$ random variables 
$X_{0, k(1)}, X_{0, k(2)},...,X_{0, k(N)}$ take the value $1$. 
Consider for each $k(l)$, $l=1,2,...,N$ ,  the random variables
 $X_{k(l), k(l)+1}$, $X_{k(l), k(l)+2},...,X_{k(l), m}$. 
Then $\sum_{j=k(l)+1}^m X_{ij}$ with $i=k(l)$   is approximately 
distributed by the Poisson distribution 
 $\frac{(c (m-k(l))^k}{k!} e^{-c(m-k(l))}$. 
 We define the Poisson  generation of intervals to  
 define  the random sequential  generation of intervals   for the cascade model as follows.

{\bf Poisson  generation of intervals}  is defined  
for  the interval $[0,y]$ to  take a random variable $N(y)$ distributed by 
the Poisson distribution with the parameter $c\, y$  as 
\begin{eqnarray*}
P(N(y)=k)=\frac{1}{k!}(c\,y )^ke^{-c\,y }
\end{eqnarray*}
to 
generate the intervals $[0,X_1(y)],[0, X_2(y)],...,[0,X_{N(y)}(y)]$,  where
 $X_i (y)$  is  distributed uniformly at random on the interval 
$[0,y]$, mutually independently, for each $i$. 

At step 1 we apply the Poisson  generation to the interval $[0,x]$.   
At step $1 <j$, for each interval 
$[0,y]$ 
generated at the step $j-1$ 
 apply the Poisson  generation to  the interval   independently from other intervals and 
independently from the previously generated intervals.  
Each interval which does not generate 
any interval at step $j$ does not generate any interval after the step $j$.  
 We continue 
the steps as long as we have  at least one generated  interval.   

We  also define the random sequential  generation of intervals with an exponentially distributed starting point as follows. At step 0 we 
generate the interval $[0,x-Z]$ where $Z$  is   distributed  by the 
density $e^{-c\, z}$.   If  ($[0,x-Z]\not \subset  [0,x] $), we finish and  stop  to make the next step.  
 If  ($[0,x-Z] \subset  [0,x] $),  we proceed to step 1.
At step 1 we apply the Poisson  generation to the interval $[0,x-Z]$.   At step $1 <j$,  for each interval 
generated at the step $j-1$ 
 apply the Poisson  generation to  the interval   independently from other intervals and 
independently from the previously generated intervals.  
Each interval which does not generate 
any interval at step $j$ does not generate any interval after the step $j$.  
 We continue 
the steps as long as we have  at least one generated  interval. 
 In the original  cascade model of  food webs, 
 the value $c\, m$ is assumed 
to be a  constant.   Assuming $c$ is a constant independent from $m$,  the  asymptotic behavior 
of the  longest chain 
is  of  mathematical interest, which  may have  applications  for example to task graphs for  
parallel processing in computer science (Newman (1992)). As shown in sections 2, 3,  4, and 5,  the random sequential  generation of intervals 
 helps to understand the asymptotic  length  of the longest chain 
of the cascade model.

\vspace{0.5cm}

\section{Random sequential  generation by the Poisson distribution \label{Poisson}}   
For each stopped interval generated by the above procedure, 
we count the number  of   steps  to  get  the stopped interval.   
Let $K(x,a)$ be the number of intervals which 
take $a$ steps until  stopping. 
Let $L(x, a)$ be the expectation of $K(x,a)$. 
We get the asymptotic behavior of $L(x,a)$ as in the case of random sequential 
bisection 
for the binary search tree 
(Sibuya and Itoh (1987)).

We have 
\begin{eqnarray}
L(x,0)=e^{-cx}.
\end{eqnarray}
For $1 \leq a$, 
we have,
\begin{eqnarray}
 L(x, a)\nonumber &=&  \frac{1}{x} \int _0^x \sum _{k=1}^{\infty} \frac{1}{k!}(c x)^k e^{-cx} 
k L(y, a-1)\,dy,
\\\nonumber
&=&   \frac{1}{x}  \int _0^x \sum _{k=1}^{\infty} \frac{1}{k!}(c x)^k e^{-cx} k L(y, a-1) \,dy,
\\\nonumber
&=&  cx \frac{1}{x} \int _0^x \sum _{k=0}^{\infty} \frac{1}{k!}(c x)^k  e^{-cx}  L(y, a-1)\,dy,\\
&=&    c  \int _0^x  L(y, a-1)\,\,dy. \label{eq:la}
\end{eqnarray} 
Hence for $a=1$, we have 
\begin{eqnarray}
 L(x, 1)\nonumber
&=&    c  \int _0^x  L(y,  0)\,\,dy, \\\nonumber
&=&    c \int _0^x e^{-cy} \,\,dy. 
\end{eqnarray} 
We have finally
\begin{eqnarray}
 L(x, a)\nonumber
 &=&  c \int _{0}^{x} c \int _{0}^{x_{a-1}}  ...   c \int _{0}^{x_1}  c^{-cx_1} \,dx_1\,dx_2\,...dx_{a-1}\\
&=&  \sum _{j=0}^{\infty} (-1)^j \frac {(c\,x)^{a+j}}{(a+j)!}.
\end{eqnarray} 
Hence we have 
\begin{eqnarray}
L(x,a)+L(x,a+1)= \frac {(c\,x)^{a}}{a!}.
\end{eqnarray}
Put $a=k \, x$, and  
apply the Stirling formula 
$n!\sim  \sqrt {2\pi}n^{n+1/2}e^{-n}$, 
then  
\begin{eqnarray}
L(x,a)+L(x,a+1)= \frac {(c\,x)^{k\, x}}{(k\, x )!}  \sim  
\frac {(c\,x)^{k \, x}}{\sqrt {2\pi}(k\,  x)^{k\,  x +1/2}e^{-k\,  x}
}  .
\end{eqnarray}
As
\begin{eqnarray}
\frac {(c\,x)^{k \, x}}{\sqrt {2\pi}(k\,  x)^{k\,  x +1/2}e^{-k\,  x}}
=\frac{1}{\sqrt {2\pi k\,x}}(\frac{c}{k})^{k\,x} e^{k\,x},
\end{eqnarray}
we have the following theorem, which corresponds to the stronger result by Newman (1992) for the cascade model.
\vspace{0.2cm}

\noindent
{\bf Theorem 1}

\noindent 
(i) 
\begin{eqnarray}
L(x,a)+L(x,a+1)= \frac {(c\,x)^{a}}{a!}.
\end{eqnarray}
(ii) Put $a=k \, x$, for $x\rightarrow \infty$,  if $e\,c <k$,   
\begin{eqnarray}
L(x,a)+L(x,a+1)\rightarrow  0,
\end{eqnarray}
if $k <e\,c$,
\begin{eqnarray}
L(x,a)+L(x,a+1)\rightarrow  \infty.
\end{eqnarray}


\section{The  chain length  with an exponentially distributed starting point}
The expected  chain length for the case that  the starting interval is $[0,x-u]$ ($\subset  [0,x] $)
 where $u$ is   distributed  by the 
density $c\, e^{-c\, u}$ is given by  
\begin{eqnarray}
M(x,a)\nonumber&=&\int _0^x  L(x-u,a) c\, e^{-c\, u}  d\,u.  
\end{eqnarray}

\begin{eqnarray}
M(x,a)+M(x,a+1)\nonumber&=&\int _0^x(L(x-u,a)+L(x-u,a+1))c\,e^{-c\, u} d\,u\\& =&\int _0^x 
\frac {(c\,(x-u))^{a}}{a!}c\, e^{-c\, u}  d\,u.  
\end{eqnarray}
\begin{eqnarray}
 M(x, a)+M(x,a+1)
&=& (-1)^a e^{-c\, x}  +  \sum _{j=0}^{a} (-1)^j \frac {(c\,x)^{a-j}}{(a-j)!}.
\end{eqnarray}
 Hence  we have Theorem 2 in the same way to Theorem 1 by using   
the Stirling formula.
\vspace{0.2cm}

\noindent
{\bf Theorem 2} 

\noindent 
(i) 
\begin{eqnarray}
M(x,a)+2\, M(x,a+1)+M(x,a+2)  
= \frac {(c\,x)^{a+1}}{(a+1)!}. \label{eq:m3}
\end{eqnarray}

\noindent
(ii) Put $a=k \, x$,  
for $x\rightarrow \infty$,  if $e\,c <k$,   
\begin{eqnarray}
M(x,a)+2\, M(x,a+1)+M(x,a+2)  
\rightarrow  0,
\end{eqnarray}
if $k <e\,c$,
\begin{eqnarray}
M(x,a)+2\, M(x,a+1)+M(x,a+2)  
\rightarrow  \infty.
\end{eqnarray}

\section{On Cascade model}
Putting  $m=S-1$,  the expected number of chains with  length $n$ on the cascade model (Cohen and Newman (1986))  
is given by  
\begin{eqnarray}
E(C_n)=p^n \sum _{k=n}^{S-1} (S-k) {k-1 \choose n-1} q^{S-k-1}.
\end{eqnarray}
By using this formula,  we can extend  equation  (\ref{eq:m3})  to the  
cascade model. The following theorem  will show that our  Poisson interval tree is  
 a natural continuous model 
for the cascade model of  food webs and  gives an elementary  approach to the theorems for the cascade model
(Newman (1992)) on  the expected length of a chain and the asymptotic length of the longest chain.

\vspace{0.2cm}

\noindent 
{\bf Theorem 3.} 

\noindent 
(i)  
\begin{eqnarray}
E(C_n)+2\,E(C_{n+1})+E(C_{n+2})    ={S \choose n+1} p^n.\label{eq:t}
\end{eqnarray}

\noindent
(ii) Put $n=k \, S\,p$,  
for $S\,p \rightarrow \infty$,  if $e\, <k$,   
\begin{eqnarray}
E(C_n)+2\,E(C_{n+1})+E(C_{n+2}) \rightarrow  0,
\end{eqnarray}
if $k <e\,$,
\begin{eqnarray}
E(C_n)+2\,E(C_{n+1})+E(C_{n+2}) \rightarrow  \infty.
\end{eqnarray}

\noindent 
{\bf Proof.}
By using 
\begin{eqnarray}
E(\frac{C_n}{p^n})= \sum _{k=n}^{S-1} (S-k) {k-1 \choose n-1} q^{S-k-1},
\end{eqnarray}

\begin{eqnarray}
E(\frac{C_{n+1}}{p^n}) \nonumber 
&=& \sum _{k=n+1}^{S-1}(1-q)  (S-k) {k-1 \choose n} q^{S-k-1}\\\nonumber
 &=& \sum _{k=n+1}^{S-1}  (S-k) {k-1 \choose n} q^{S-k-1}\\
&-& \sum _{k=n+1}^{S-1}(S-k) {k-1 \choose n} q^{S-k},
\end{eqnarray}
and 
\begin{eqnarray}
E(\frac{C_{n+2}}{p^n}) \nonumber &=&  \sum _{k=n+2}^{S-1} (1-q)^2 (S-k) {k-1 \choose n+1} q^{S-k-1}\\\nonumber 
&= & \sum _{k=n+2}^{S-1}  (S-k) {k-1 \choose n+1} q^{S-k-1}\\\nonumber 
& -& \sum _{k=n+2}^{S-1} 2\,(S-k) {k-1 \choose n+1} q^{S-k}\\
& +& \sum _{k=n+2}^{S-1}  (S-k) {k-1 \choose n+1} q^{S-k+1}, 
\end{eqnarray}
we can easily obtain eq. (\ref{eq:t}).

 (ii) is obtained by using the Stirling formula as in the proof  of Theorem 1 and Theorem 2.

\section{On the height of the generated  tree}   
For each  stopped interval generated by the above procedure in the section, 
we count the number  of   steps $a$ to  get  the stopped interval.   Consider 
the maximum $H(x)$  of the numbers  and let us  call it the height  of the generated  tree.
For  the probability  $F(x,h)\equiv Pr(H(x)\leq h)$ of the height of  Poisson interval 
tree $H(x)$,  we have for $h=0$, 
\begin{eqnarray}
F(x,h)=e^{-cx},
\end{eqnarray}
for $1 \leq h$, 
\begin{eqnarray}
\nonumber & &F( x,h)\\\nonumber 
&=&  e^{-cx}+ \sum _{k=1}^{\infty} \frac{(c x)^k }{k!}e^{-cx}  \frac{1}{x^k}  \int _0^x \cdots  \int _0^x  F( y_1,h-1)\cdots F(  y_k, h-1)\,dy_1\cdots dy_k,
\\\nonumber
&=&  e^{-cx}  \sum _{k=0}^{\infty} \frac{1}{k!} ( c  \int _0^x  F( y, h-1)\,dy)^k
\\
&=&  e^{-cx}  e^{( c  \int _0^x  F(y, h-1)\,dy)}. 
\label{eq:height}
\end{eqnarray} 
We can sequentially integrate   at each step and obtain  $F(x,h)$ starting from 
 $F(x,0)=e^{-cx}$. The expansion at $x=0$ is obtained 
 by Mathematica up to $(c x) ^{24}$ for example for  $h=0, 1, 2, ..., 7$, which  gives reasonable numerical  values of $F(x, h)- F(x,h-1)$ when $ c x  \leq 2.1$, 
 while it does not give reasonable numerical values for $2.2 \leq c x$, 
 For $4\leq h$ the first 
 five terms of  the probability $F(x, h)- F(x,h-1)$ that the 
the height is $h$ is given by 
 \begin{eqnarray}
\frac{(cx)^h}{h !}-2\frac{(cx)^{h+1}}{(h+1) !}-\frac{(cx)^{h+2}}{(h+2) !}+\frac{(cx)^{h+3}}{(h+3) !}-8\frac{(cx)^{h+4}}{(h+4) !}
\label{eq:height2}. 
\end{eqnarray} 
\vspace{0.5cm}

\noindent 
{\bf Acknowledgment}  I thank  Joel E. Cohen for his helpful  comments and  discussion. 
This research was supported in part by US National Science Foundation 
Grant   DMS 0443803 to Rockefeller University.


\begin{thebibliography}{11}
\bibitem{c}
Cohen, J. E.(1990).
 A stochastic theory of community food webs. VI. 
Heterogeneous alternatives to the cascade model,
 Theoretical Population Biology,  37, 55--90.

\bibitem{cbn}
Cohen, J.  E., Briand, F.  and Newman, C.  M. (1990).  
 Community food webs: Data and Theory, Springer-Verlag,  New York.

\bibitem{cn}
Cohen, J.  E. and Newman, C.  M. (1985).  
 A stochastic theory of community food webs.  I. 
Models and aggregated data, 
 Proceedings of the   Royal   Society  ( London),  B 224, 421--448.

\bibitem{cn}
Cohen, J.  E. and Newman, C.  M. (1986).  
 A stochastic theory of community food webs. I. 
Models and aggregated data, 
 Proceedings of the   Royal   Society  ( London),   B 228, 355--377.

\bibitem{d} Devroye, L. (1986). A note on the height of binary search trees.  
   Journal of the Association of Computing Machinery, 
    33, 489--498.

\bibitem{dk} Durrett, R. and Kesten, H. (1990).  The critical parameter of the connectedness of some 
random graphs,  
   A Tribute to Paul Erd\"os,  161--176, Cambridge University Press.
   
\bibitem{di} Dutour Sikiric, M. and Itoh, Y. (2011).  Random sequential packing of cubes,
World Scientific.

\bibitem{er} Erd\"os, P. and   R\'enyi, A.  (1960). On the evolution of random graphs,
    Publ. Math. Inst. Hungar. Acad. Sci. 5, 17-61. 

\bibitem{fo} Flajolet, P. and Odlyzko, A. (1982). The average height of binary tree and othe simple tree, 
Journal of Computer and System Sciences,   25, 171-213. 

\bibitem{jn} Janson, S. and Neininger, R. (2008). 
The size of random fragmentation trees, Probability Theory and Related Fields,
 142, 399-442.

\bibitem{mp} Mahmoud, H. and Pittel, B. (1984). On the most probable shape of a search tree grown from 
random permutations. SIAM Journal on Algebraic and Discrete methods,  1,  69-81. 

\bibitem{n}
Newman, C. M. (1992). 
 Chain Lengths in Certain Random Directed Graphs, 
 Random Structures and Algorithms,   3, 243--253.
 
\bibitem{rb} Robson, J. M. (1979).
   The height of binary search trees, 
 The Australian Computer Journal, 
 11, 151-153.

\bibitem{s} Shepp, L. A. \ (1989). 
   Connectedness of certain random graphs,
   Israel Journal of  Mathematics,
    67, 23--33.

\bibitem{si} Sibuya, M.\ and Itoh, Y.\ (1987). 
   Random sequential bisection and its associated binary tree, 
    Annals of the Institute of Statistical Mathematics, 
    39, 69--84.
\end{thebibliography}
\end{document}